\documentclass[twocolumn,superscriptaddress,showpacs,pra]{revtex4}
\usepackage{epsfig}
\usepackage{epstopdf}
\usepackage{delarray}
\usepackage{amsmath, amssymb}
\usepackage{bm}
\usepackage{graphicx}
\usepackage{placeins}
\usepackage{color}

\begin{document}
\title{Self-localized Solitons of a q-Deformed Quantum System}	

\author{Cihan Bay\i nd\i r}
\affiliation{\.{I}stanbul Technical University, Engineering Faculty, 34469 Maslak, \.{I}stanbul, Turkey.\\  
Bo\u{g}azi\c{c}i University, Engineering Faculty, 34342 Bebek, \.{I}stanbul, Turkey. \\
CERN, CH-1211 Geneva 23, Switzerland.}
\email{cbayindir@itu.edu.tr}

\author{Azmi Ali Altintas}
\affiliation{Department of Electrical Engineering, \.{I}stanbul Okan University, 34959 Tuzla, \.{I}stanbul, Turkey}
\email{altintas.azmiali@gmail.com}

\author{Fatih Ozaydin}
\affiliation{Institute for International Strategy, Tokyo International University, 1-13-1 Matoba-kita, Kawagoe, Saitama, 350-1197, Japan}
\email{fatih@tiu.ac.jp}

\begin{abstract}
Beyond a pure mathematical interest, q-deformation is promising for the modeling and interpretation of various physical phenomena. In this paper, we numerically investigate the existence and properties of the self-localized soliton solutions of the nonlinear Schr\"{o}dinger equation (NLSE) with a q-deformed Rosen-Morse potential. By implementing a Petviashvili method (PM), we obtain the self-localized one and two soliton solutions of the NLSE with a q-deformed Rosen-Morse potential. In order to investigate the temporal behavior and stabilities of these solitons, we implement a Fourier spectral method with a  $4^{th}$ order Runge-Kutta time integrator. We observe that the self-localized one and two solitons are stable and remain bounded with a pulsating behavior and minor changes in the sidelobes of the soliton waveform. Additionally, we investigate the stability and robustness of these solitons under noisy perturbations. A sinusoidal monochromatic wave field modeled within the frame of the NLSE with a q-deformed Rosen-Morse potential turns into a chaotic wavefield and exhibits rogue oscillations due to modulation instability triggered by noise, however, the self-localized solitons of the NLSE with a q-deformed Rosen-Morse potential are stable and robust under the effect of noise. We also show that soliton profiles can be reconstructed after a denoising process performed using a Savitzky-Golay filter.
\end{abstract}


\maketitle
\section{Introduction}

Quantum harmonic oscillator (QHO) is widely accepted and commonly implemented as a model to study the atomic and molecular vibrations \cite{Schrodinger, Griffiths, Liboff, Pauli, Messiah}. The most common form of the QHO involves linear Schr\"{o}dinger equation and a potential term in the form of a parabolic trap. This form of the QHO is one of the rare solvable models in quantum mechanics and admits solutions in the form of Hermite polynomials \cite{Schrodinger, Griffiths, Liboff, Pauli, Messiah}. There are many possible extensions of the QHO. While it is possible to extend it to higher dimensions \cite{Pauli, Messiah}, some researchers have proposed and investigated various forms of the nonlinear quantum harmonic oscillator (NQHO) \cite{Kivshar, Cihan2020PhysA, Carinena2007, Zheng, Ranada, SchulzeHalberg2012, SchulzeHalberg2013}. Some of the literature is devoted to the investigation of the QHO under significantly nonlinear electric fields which eventually leads to various forms of the nonlinear Schr\"{o}dinger equation (NLSE) \cite{Kivshar,Cihan2020PhysA}. Similar equations of NLSE type are also studied to model Bose-Einstein condensation and soliton tunneling where the potential function is parabolic type \cite{PerezGarcia,Atre,Liang,Rajendran,ZhongBelic}. Some other researchers have investigated various forms of potentials behaving nonlinearly and their effect on the quantum harmonic oscillations \cite{Carinena2007, SchulzeHalberg2012, SchulzeHalberg2013}.

On the other hand, the idea of the quantum group was developed for the quantization of inverse scattering theory. In order to solve the 1+1 dimensional quantum integrable systems, this quantum inverse scattering theory can be used \cite{Faddeev}. From a mathematical perspective, a quantum group is basically a Hopf algebra, and the algebraic structure must satisfy the Hopf algebra axioms. There are many techniques to construct a quantum group, such as  Woronowicz's method \cite{Wor}, Manin's approach \cite{Manin} and Drinfeld's approach \cite{Drinfeld}. While Woronowicz's approach utilizes non-commutative multiplication, in Manin's approach a quantum group is constructed by making linear transformation on the quantum plane.  Manin's approach can also be used to construct the inhomogeneous invariance quantum groups \cite{invariance1, invariance2, invariance3, invariance4}. 

In Drinfeld's approach, a deformation parameter q is defined on Lie algebra \cite{Drinfeld}. Although a Lie algebra is a subject of mathematics, q deformation has deep meaning in physics. The idea of deformation was applied to particle algebras and the first examples of deformed particle algebra is the generalization of boson algebra \cite{arik, arik2}. Macfarlane \cite{mac} and Biedenharn \cite{biedenharn, Biedenharnbook} showed the relation between q-deformed boson algebra and $SU_q(2)$.

Another interesting application of q-deformation is thermodynamics. The generalization of Boltzmann-Gibss entropy is constructed by Tsallis \cite{Tsallis}. The search for thermodynamic properties of q-deformed systems, such as those investigated in \cite{Lavagno2002PRE, Algin, Algin2, Guha}, is still an active area of research. Also, as the quantum thermodynamics becomes more popular \cite{Scully2003Science, Turkpence2016PRE, Hardal2018PRE, Dag2019JPCC, Tuncer2019QIP}, the q-deformation gets another significant role, that is the relationship between entropy and information theory implies a new research area for q-deformed systems. 

The q-deformation has found several application areas in quantum information as well. Hasegawa studied quantum Fisher information and q-deformed relative entropies \cite{Hasegawa}. Two- and three-particle logic gates were derived for qubit and qutrit states based on q-deformation \cite{Gangopadhyay,Altintas2014QIP,Altintas2020QIP}. 
Nonclassical properties of noncommutative q-deformed cat states were analyzed by Sanjib \cite{Sanjib2015PRD}, and Berrada and Eleuch \cite{BerradaEleuch2019PRD}, who also studied entanglement of nonlinear systems in the context of q-deformation in q-Heisenberg Weyl algebra \cite{Berrada2012QIP}.

The q-deformation is also used to fit the theoretical results to the experimental ones in nuclear and atomic physics \cite{bonatsos, bonatsos2, georgieva, hammad, Jarvis2016JPA, jafarizadeh}, and Sviratcheva et al. revealed the significant role of q-deformation in understanding higher-order effects in the many-body interactions \cite{sviratcheva}. Various dynamic equations of NLSE type with q-deformed potentials have been proposed to model q-deformed processes in nuclear and atomic physics (see i.e. \cite{Falaye2013CPB, Pradeep, Filippov} and the references therein), and very recently Nutku et al. studied the complexity of q-deformed quantum harmonic oscillator \cite{Nutku2019}.

In this paper, we numerically investigate the self-localized soliton solutions of the NLSE with a q-deformed potential. The q-deformed potential we investigate is of Rosen-Morse type which is commonly used in the literature to model the q-deformation \cite{Falaye2013CPB}. In order to construct soliton solutions of the NLSE with a q-deformed Rosen-Morse potential starting from arbitrary initial conditions, we implement Petviashvili's method (PM) \cite{Petviashvili}. We discuss the effects of q-deformed Rosen-Morse potential on the characteristics of the self-localized one and two soliton solutions of the NLSE. It is a known fact that the self-localized solitons may be stable or unstable. Thus, in order to investigate the stability characteristics of those solitons, we calculate the soliton eigenvalue vs power graphs in order to assess the Vakhitov-Kolokolov condition which is one of the necessary conditions for the linear soliton stability. Then, by implementing a Fourier spectral method with a $4^{th}$ order Runge-Kutta time integrator, we investigate the nonlinear stability and temporal dynamics of the solitons of the NLSE with a q-deformed Rosen-Morse potential. We observe that self-localized one and two soliton solutions of this model are stable and the main lobe of the soliton waveform is well-conserved during temporal evolution and only minor sidelobe changes are observed which have a pulsating behavior. We display the time vs soliton power graphs which are necessary to control the stabilities of those solitons. Additionally, we investigate the stabilities and robustness of these solitons under noise by imposing a white noise term to the self-localized solitons during time-stepping. When a sinusoidal monochromatic wave within the frame of the NLSE with a q-deformed Rosen-Morse potential is subjected to noise, the modulation instability turns the monochromatic wavefield into a chaotic one which exhibits rogue harmonic oscillations, which can be basically defined as high and unexpected waves having a standing behavior. We discuss the effect of q-deformation on the statistics of these chaotic wavefields. However, contrary to this result, we observe that the self-localized solitons of the NLSE with a q-deformed Rosen-Morse potential remain stable and robust under the effect of noise. We also show that the noise can be removed and the main soliton profiles can be reconstructed after a denoising process performed using a Savitzky-Golay filter. We discuss our findings and comment on the applicability of our results.

\section{Mathematical Formulation and Discussion}

Various forms of the NLSE and their extensions with different types of potential functions are commonly used as models in the fields including but are not limited quantum mechanics \cite{Kivshar, Cihan2020PhysA, BayZeno}, nonlinear optics \cite{Fibich, Akhmediev2009b, Akhmediev2009a, Akhmediev2011, BayPRE1, BayPRE2}, hydrodynamics \cite{Kharif, BayPLA}. Following these studies, the non-dimensional NLSE equation with a potential term can be written as
\begin{equation}
i\psi_t +  \psi_{xx}+\sigma \left| \psi \right|^2 \psi-V(x)\psi=0
\label{eq01}
\end{equation}

\begin{figure}[b!]
	\begin{center}
		\hspace*{-0.7cm}
		\includegraphics[width=4.0in]{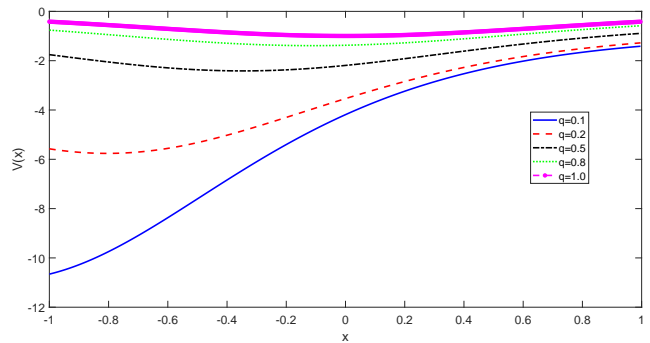}
	\end{center}
	\caption{\small q-deformed Rosen-Morse type potential for various q values.}
	\label{fig1}
\end{figure}

\noindent where $\psi$ denotes the complex wavefunction and $t, x$ denote the temporal and spatial parameters, respectively. Here, $i$ is the imaginary unity and $\sigma$ is a real constant which controls the nonlinearity. Throughout this study, this parameter is select as $\sigma=1$. In Eq.~(\ref{eq01}), $V(x)$ denotes the potential function. Some of the commonly used potentials are parabolic traps, photorefractive and saturable photorefractive potentials, Coulomb potentials, just to name a few. In this paper, we investigate the of q-deformed potentials and investigate the self-localized soliton dynamics of the NLSE with a q-deformed potential. There are some known solutions of the different nonlinear systems with q-deformed potential such as those given by Falaye et al. \cite{Falaye2013CPB}, and Eleuch \cite{Eleuch}, however, they are valid either for steady state or for different models. With this motivation, we consider the unsteady NLSE with the q-deformed Rosen-Morse potential in this study. The q-deformed Rosen-Morse potential (see i.e. \cite{Falaye2013CPB} and the references therein) can be written as: $V(x)=A\text{sech}_q^2(\alpha x)+ B\text{tanh}_q(\alpha x)$ where q-deformed hyperbolic functions are defined as $\text{sech}_q(x)=\frac{2}{e^{x} + q e^{-x}}$ and $\text{tanh}_q(x) =\frac{\text{sinh}_q(x)}{\text{cosh}_q(x)}= \frac{e^{x} - q e^{-x}}{e^{x} + q e^{-x}}$ \cite{Falaye2013CPB}. The shape of the q-deformed Rosen-Morse potential is displayed in Fig.~\ref{fig1} for various values of q.

The usage and benefits of using this type of q-deformed potential is discussed in the preceding section of this paper by referencing the relevant literature. In this paper, we investigate the self-localized soliton solutions of the NLSE given by Eq.~(\ref{eq01}) with a q-deformed Rosen-Morse potential function displayed in Fig.~(\ref{fig1}).

It is possible to obtain the self-localized solutions of different nonlinear systems using numerous computational and analytical techniques \cite{Petviashvili, Ablowitz, Yang, Bay_CSRM}. Some of these methods are shooting, self-consistency and relaxation methods \cite{Petviashvili, Ablowitz, Yang, Bay_CSRM}. Among these methods, the Petviashvili method (PM) is a popular choice \cite{Petviashvili}. In PM, the governing nonlinear equation is transformed into wavenumber domain by utilizing Fourier transforms and iterations of the scheme are performed until a convergence criterion is achieved. The convergence criterion depends on the degree of nonlinearity \cite{Petviashvili, Ablowitz, Yang, Bay_CSRM}. This method, which is known as PM, was initially proposed by Petviashvili and implemented to obtain the soliton solutions of the Kadomtsev-Petviashvili (KP) equation in \cite{Petviashvili}. Later, PM was commonly used to analyze various nonlinear physical phenomena such as dark and gray solitons and lattice vortices, just to name a few \cite{Ablowitz, Yang, BayRINP}. Since PM only works well for nonlinearities with fixed homogeneity and its extension which is known as the spectral renormalization method (SRM) is proposed to deal with other types of nonlinearities \cite{Fibich, Ablowitz}. Later, in order to obtain self-localized solitons when there are missing spectral data, another extension which can be named as compressive spectral renormalization method (CSRM) is proposed by one of us \cite{Bay_CSRM}.

The PM transforms the governing nonlinear equation into wavenumber space using FFts and iteration are performed in the Fourier space until convergence of the scheme is achieved. The convergence criterion is basically an energy conservation principle \cite{Petviashvili, Ablowitz}. We refer the reader to \cite{Fibich, Petviashvili, Ablowitz} (and the references therein) for a more comprehensive discussion of the PM, its extensions and applications.

In this section, we apply the PM to the NLSE with a q-deformed Rosen-Morse type potential to obtain its self-localized soliton solutions. With this purpose, we rewrite the NLSE given in Eq.~(\ref{eq01}) as

\begin{equation}
i\psi_t +  \psi_{xx} -V(x)\psi+ N(\left| \psi \right|^2) \psi =0
\label{eq02}
\end{equation}
where $N(\left| \psi \right|^2)=\sigma \left| \psi \right|^2$ is the nonlinear term. Pluging in the ansatz, $\psi(x,t)=\eta(x,\mu) \textnormal{exp}(i\mu t)$ the NLSE with q-deformed Rosen-Morse potential becomes
\begin{equation}
-\mu \eta +  \eta_{xx} -V(x)\eta+ N(\left| \eta \right|^2) \eta =0
\label{eq03}
\end{equation}
where $\mu$ denotes the soliton eigenvalue. Usin the definition of the 1D Fourier transform, the Fourier transform of the parameter $\eta$ can be obtained as
\begin{equation}
\widehat{\eta} (k)=F[\eta(x)] = \int_{-\infty}^{+\infty} \eta(x) \exp[i(kx)]dx
\label{eq04}
\end{equation}

It is possible to derive the spectral iteration equation by taking the 1D Fourier transform of Eq.~(\ref{eq03}). However, it is known that the iteration equation which can be derived this way would lead to a singularity. In order to avoid singularity of the iteration scheme, we add and substract a $p \eta$ term from Eq.~(\ref{eq03}) where $p$ denotes a real number with $p>0$ \cite{Petviashvili, Ablowitz}. This parameter is selected as $p=1$ throughout this study. After this operation, the 1D Fourier transform of Eq.~(\ref{eq03}) gives	
\begin{equation}
\widehat{\eta} (k)=\frac{(p+| \mu|)\widehat{\eta}}{p+\left| k \right|^2} -\frac{F[V \eta]-F \left[ N(\left| \eta \right|^2) \eta \right]}{p+\left| k \right|^2}
\label{eq05}
\end{equation}
which is our iteration scheme for the NLSE with a q-deformed Rosen-Morse potential. In order to obtain a convergence criterion for the iterations, one can introduce a new parameter $\eta(x)=\gamma \xi(x)$, and denote its Fourier transform as $\widehat{\eta}(k)=\gamma \widehat{\xi}(k)$. Using this substitution, the iteration formula given for the NLSE with a q-deformed Rosen-Morse potential becomes
\begin{equation}
\begin{split}
\widehat{\xi}_{j+1} (k) =\frac{(p+| \mu|)}{p+\left| k \right|^2}\widehat{\xi_j}-\frac{F[V \xi_j]}{p+\left| k \right|^2} + & \frac{F\left[\sigma |\gamma_j|^2|\xi_j|^2 \xi_j \right] }{p+\left| k \right|^2} \\
& =R_{\gamma_j}[\widehat{\xi}_j (k)]
\label{eq06}
\end{split}
\end{equation}
The convergence criteria for PM can be attained by multiplying both sides of Eq.~(\ref{eq06}) with $\widehat{\xi}^*(k)$ and integrating to evaluate the total energy. After these operations, the normalization constraint becomes
\begin{equation}
\int_{-\infty}^{+\infty} \left|\widehat{\xi} (k)\right|^2 dk= \int_{-\infty}^{+\infty} \widehat{\xi}^* (k) R_{\gamma}[\widehat{\xi} (k)]dk  
\label{eq07}
\end{equation}
The constraint given in Eq.~(\ref{eq07}) ensures that the scheme converges to self-localized solitons of the dynamic equation under investigation. Then, starting from the initial conditions, Eq.~(\ref{eq05}) and the energy constraint given in Eq.~(\ref{eq07}) are applied iteratively to find the profile at each iteration count. Iterations are ceased after the parameter ${\gamma}$ convergences to a specified error bound. This error bound is selected to be $10^{-5}$ and used for all simulations presented in this paper.

\begin{figure}[t!]
	\begin{center}
		\hspace*{-0.7cm}
		\includegraphics[width=4.0in]{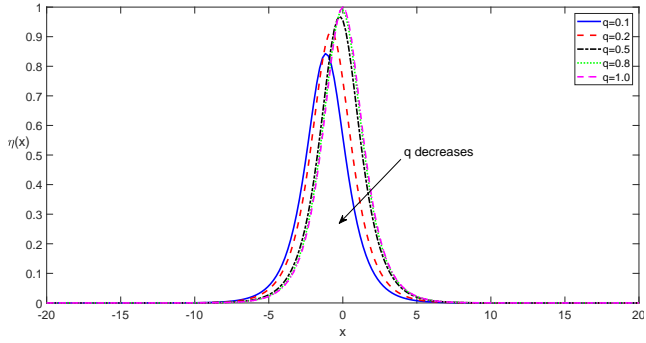}
	\end{center}
	\caption{\small Self-localized one soliton solutions of the NLSE with the q-deformed Rosen-Morse type potential for various values of q.}
	\label{fig2}
\end{figure}

\begin{figure}[t!]
	\begin{center}
		\hspace*{-0.7cm}
		\includegraphics[width=4.0in]{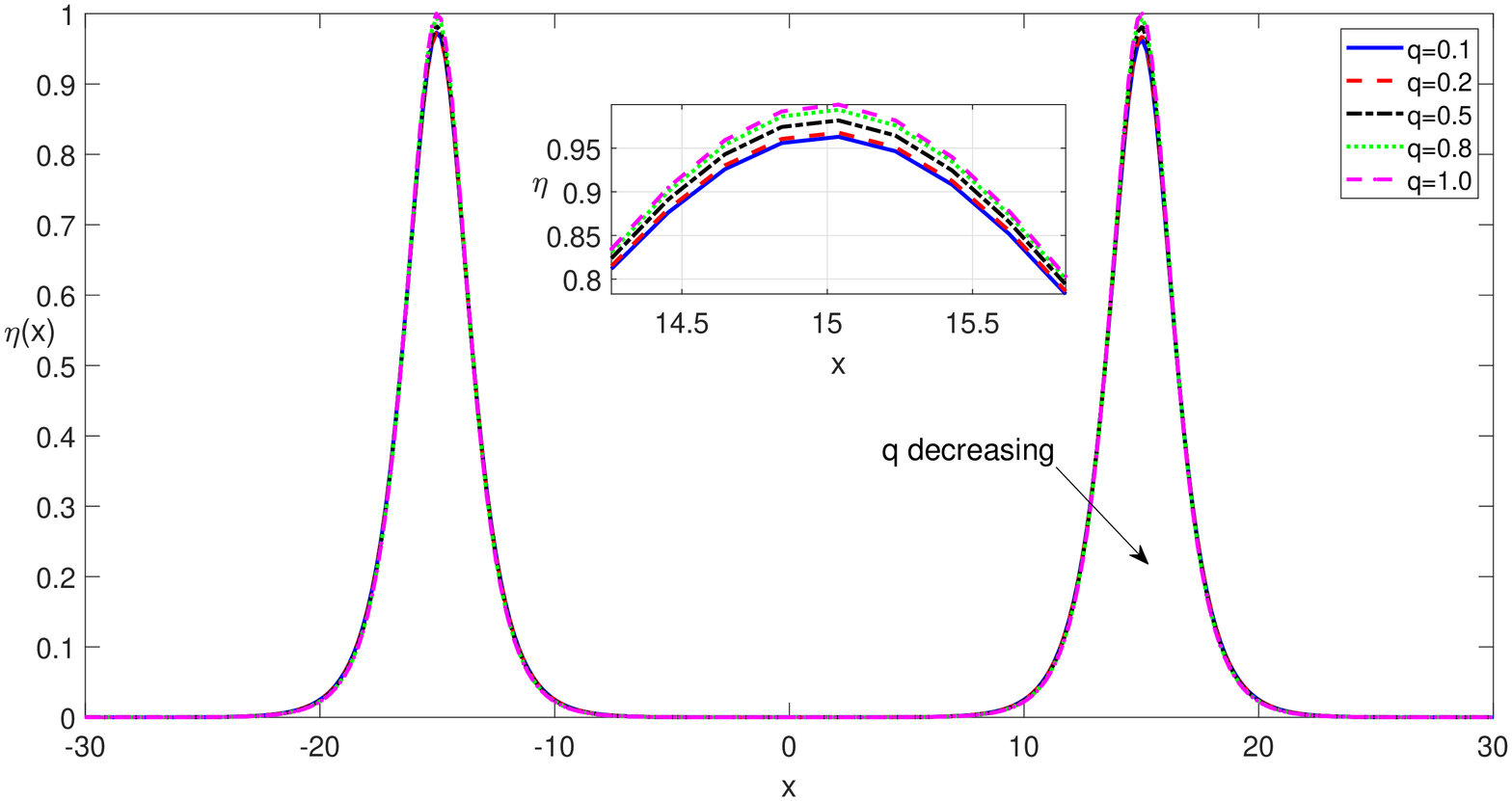}
	\end{center}
	\caption{\small Self-localized two soliton solutions of the NLSE with the q-deformed Rosen-Morse type potential for various values of q.}
	\label{fig3}
\end{figure}

Using a Gaussian in the form of $\exp{(-x^2)}$ as the initial condition, we implement the PM summarized above for various values of the deformation parameter q. The computational domain length and domain are selected to be $L=80$ and $x=[-L/2,L/2]$, respectively. The number of spectral components is selected as $N=1024$. The self-localized one soliton solutions of the NLSE with the q-deformed Rosen-Morse type potential is depicted in Fig.~\ref{fig2} for various values of deformed parameter q. In this plot, the amplitude of the self-localized soliton obtained for the non-deformed case ($q=1$) is used as the normalization amplitude. Checking Fig.~\ref{fig2}, one can observe that a decrease in the deformation parameter q leads to a scaling of the soliton amplitude. Additionally, a significant spatial shift in the location of the peak of the soliton is also observed. Checking the odd parts of the solitons, we did not observe any skewness in the soliton profile.

When the simulations are initiated using two Gaussians leading to an initial condition in the form of $\exp{(-(x-15)^2)}+\exp{(-(x+15)^2)}$, the PM converges to self-localized two soliton solutions of the NLSE with the q-deformed Rosen-Morse potentials. These two peaked solitons are displayed in Fig.~\ref{fig3} for various values of the deformation parameter q. Compared to the one soliton solutions, these two solitons do not exhibit a significant shift in the location of the peak when different values of q are used. Also, the decrease in soliton amplitudes becomes less significant. 
\begin{figure}[t!]
	\begin{center}
		\hspace*{-0.7cm}
		\includegraphics[width=4.0in]{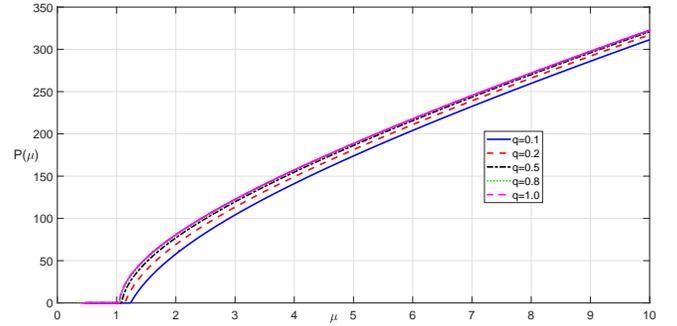}
	\end{center}
	\caption{\small Self-localized one soliton power vs soliton eigenvalue, $\mu$.}
	\label{fig4}
\end{figure}

\begin{figure}[t!]
\begin{center}
\hspace*{-0.7cm}
   \includegraphics[width=4.0in]{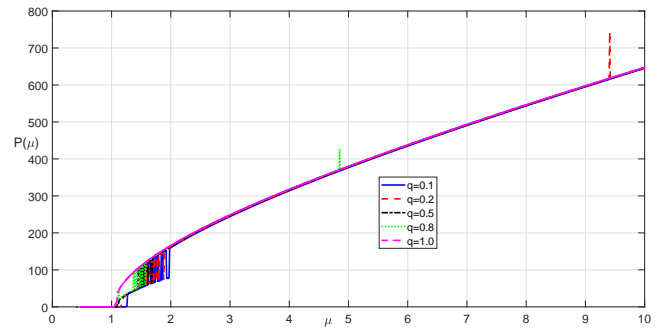}
  \end{center}
\caption{\small Self-localized two soliton power vs soliton eigenvalue, $\mu$.}
  \label{fig5}
\end{figure}

Regarding soliton analysis, one of the questions of practical importance is whether the solitons found by PM are stable or not. There are some studies which investigate the stability characteristics of solitons (see i.e. \cite{VakhitovStability, WeinsteinStability, SivanStability} and the references therein). One of the necessary, but not sufficient conditions for soliton stability is the Vakhitov-Kolokolov condition, which is also known as the slope condition \cite{VakhitovStability}. This condition states that when the positive values of the soliton eigenvalue are considered, that is for $\mu>0$, the $dP/d \mu >0$ condition should hold true \cite{VakhitovStability}. Here, $P=\int |\psi|^2dx$ is the soliton power. In order to check if this criterion is satisfied, we plot the soliton eigenvalue vs soliton power graphs for self-localized one and two solitons in Fig.~\ref{fig4} and Fig.~\ref{fig5}, respectively. For the range of soliton eigenvalues, i.e. $\mu \in [0,10]$, considered, we observe that the Vakhitov-Kolokolov condition with the exceptions of some spikes in Fig.~\ref{fig5}. Thus, the solitons eigenvalue of $\mu=2$ is used for all the simulations presented throughout this paper.
 
Although the Vakhitov-Kolokolov condition is necessary for the soliton stability, it is not sufficient. Some forms of the spectral conditions for the soliton stability are discussed in \cite{WeinsteinStability, SivanStability}. Under some circumstances, the analytical forms of the spectral condition may be derived. However, researchers generally assess the stability of solitons by observing their temporal dynamics. With this motivation, we implement a Fourier spectral scheme with $4^{th}$ order Runge-Kutta time integrator for the time stepping of the NLSE with the q-deformed Rosen-Morse potential. We begin by rewriting the NLSE given by Eq.(\ref{eq01}) as
\begin{equation}
\psi_t =i \left( \psi_{xx} -V(x) \psi + \sigma  \left|\psi \right|^2 \psi \right) =h(\psi, t,x)
\label{eq08}
\end{equation}
Here, $h(\psi, t,x)$ is a new function used to denote the right-hand-side of Eq.(\ref{eq08}). The second order spatial derivative in Eq.(\ref{eq08}) can be calculated using FFTs at each time step as
\begin{equation}
\psi_{xx} = F^{-1} \left[ -k^2 F[\psi] \right]
\label{eq09}
\end{equation}

\noindent where $F$ and $F^{-1}$ shows the forward and the inverse FFTs, respectively \cite{Canuto, Karjadi2012, Baysci, trefethen, BayMS}. In Eq.(\ref{eq09}), the parameter $k$ denotes the wavenumber vector. This wavenumber vector contains $N$ multiples of the fundamental wavenumber, $k_o=2 \pi/L$. As before, the number of spectral components and the domain length is selected as $N=1024$ and $L=80$, respectively. The selection of the number of spectral coefficients as a power of 2 ensures the efficient computations of FFTs. The nonlinear products are in the physical domain by simple multiplication. For the time stepping of Eq.(\ref{eq08}), a $4^{th}$ order Runge-Kutta scheme is implemented. Four slopes of the $4^{th}$ order Runge-Kutta at each time index, $n$, can be calculated using 
\begin{equation}
\begin{split}
& s_1=h(\psi^n, t^n, x) \\
& s_2=h(\psi^n+0.5 s_1dt, t^n+0.5dt, x) \\
& s_3=h(\psi^n+0.5 s_2dt, t^n+0.5dt, x) \\
& s_4=h(\psi^n+s_3dt, t^n+dt, x) \\
\end{split}
\label{eq10}
\end{equation}

\noindent where $dt$ is the time step. It is selected as small as $dt=5 \times 10^{-4}$ throughout this study, to prevent the blow-up of the numerical solutions. Then, the complex wavefunction and time parameter at the next time steps of evolution are be found by
\begin{equation}
\begin{split}
& \psi^{n+1}=\psi^{n}+(s_1+2s_2+2s_3+s_4)/6 \\
& t^{n+1}=t^n+dt\\
\end{split}
\label{eq11}
\end{equation}
where $n$ denotes the iteration count \cite{Canuto, Karjadi2012, Baysci, trefethen, BayMS}. Using the self-localized solitons of the NLSE with the q-deformed Rosen-Morse potential as the initial conditions, it is possible to analyze their temporal dynamics and the stability using the computational procedure summarized above.

 \begin{figure}[t!]
	\begin{center}
		\hspace*{-0.7cm}
		\includegraphics[width=4.0in]{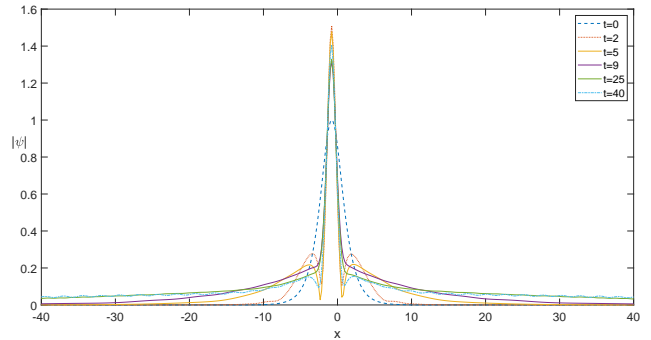}
	\end{center}
	\caption{\small Self-localized one soliton profiles at different times for $q=0.2$.}
	\label{fig6}
\end{figure}

\begin{figure}[b!]
	\begin{center}
		\hspace*{-0.7cm}
		\includegraphics[width=4.0in]{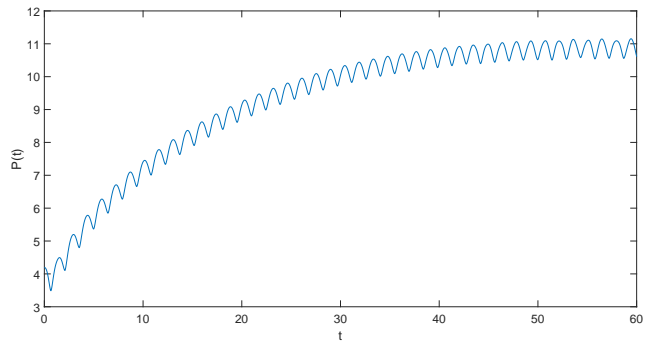}
	\end{center}
	\caption{\small Self-localized one soliton power as a function of time for $q=0.2$.}
	\label{fig7}
\end{figure}

\begin{figure}[t!]
	\begin{center}
		\hspace*{-0.7cm}
		\includegraphics[width=4.0in]{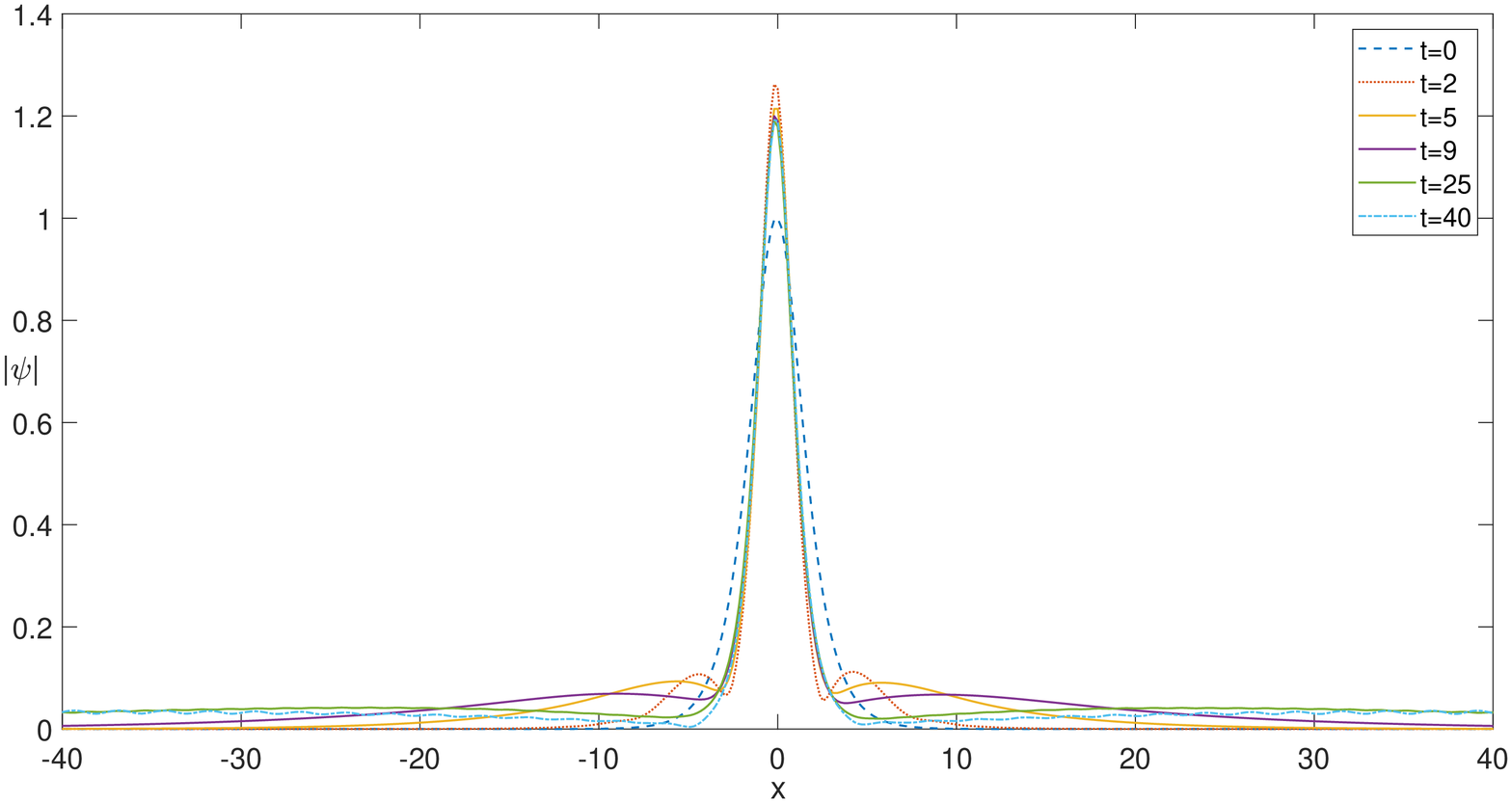}
	\end{center}
	\caption{\small Self-localized one soliton profiles at different times for $q=0.8$.}
	\label{fig8}
\end{figure}

\begin{figure}[b!]
	\begin{center}
		\hspace*{-0.7cm}
		\includegraphics[width=4.0in]{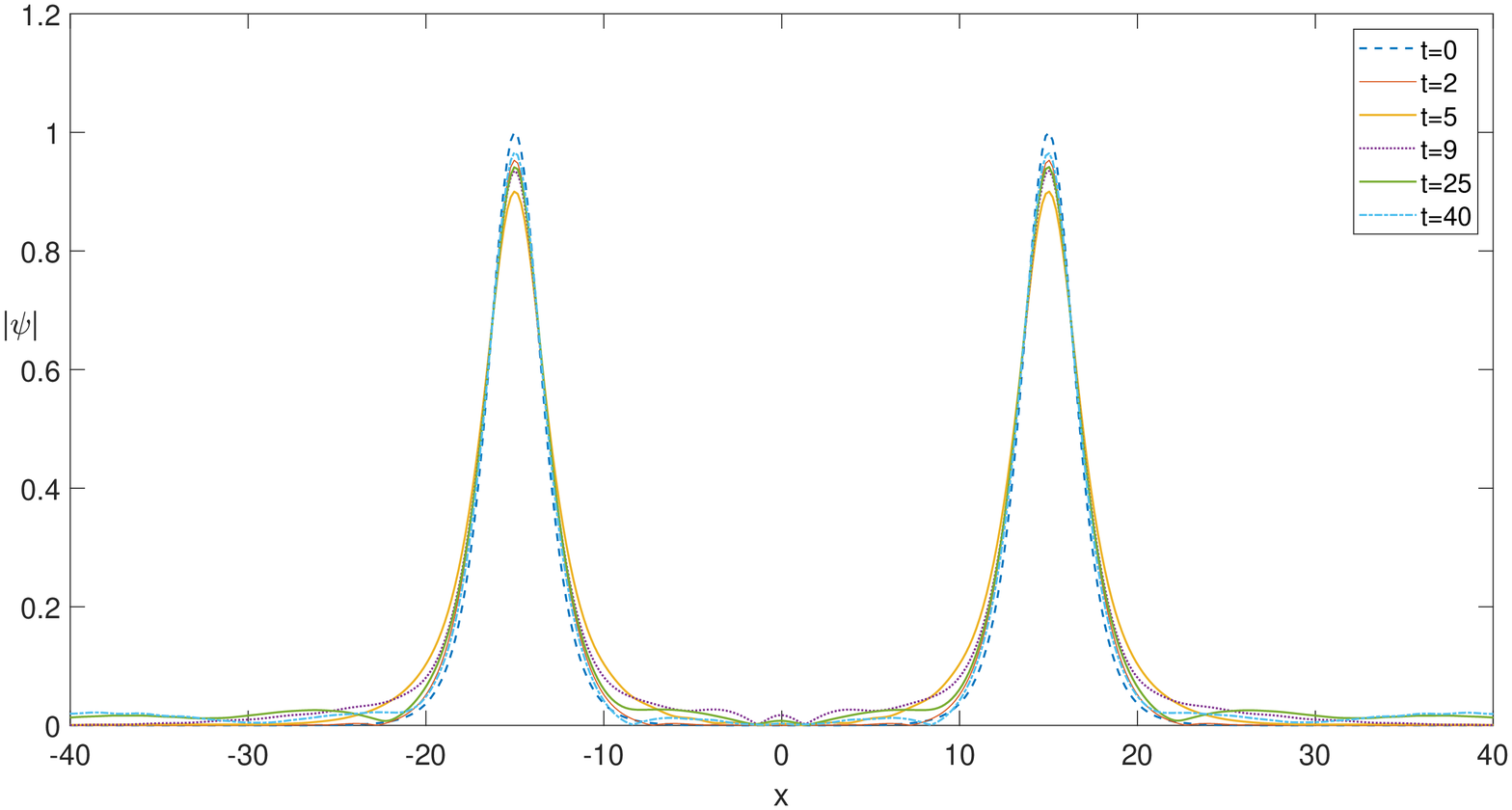}
	\end{center}
	\caption{\small Self-localized two soliton profiles at different times for $q=0.8$.}
	\label{fig9}
\end{figure}
\begin{figure}[t!]
	\begin{center}
		\hspace*{-0.7cm}
		\includegraphics[width=4.0in]{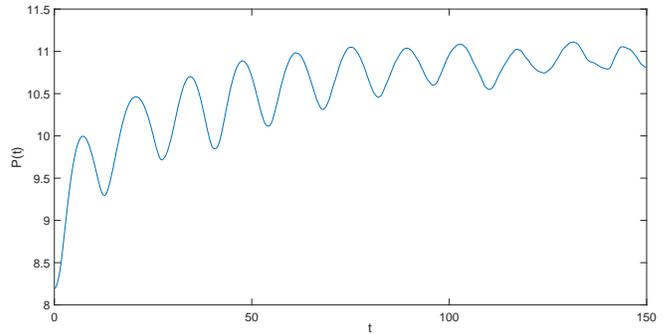}
	\end{center}
	\caption{\small Self-localized two soliton power as a function of time.}
	\label{fig10}
\end{figure}

\begin{figure}[b!]
	\begin{center}
		\hspace*{-0.7cm}
		\includegraphics[width=4.0in]{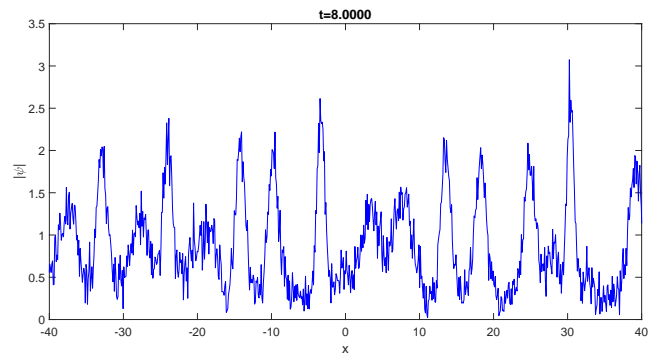}
	\end{center}
	\caption{\small A chaotic wavefield generated in the frame of the q-deformed NLSE using $q=0.8, \beta=0.2$ exhibiting rogue oscillations.}
	\label{fig11}
\end{figure}
We depict the self-localized soliton profiles obtained by this computational algorithm at different times in Fig.~\ref{fig6} for $q=0.2$ and the associated time vs soliton power graph in Fig.~\ref{fig7}.

Checking Fig.~\ref{fig6}, one can observe that during the temporal evolution main lobe of the soliton is well-preserved.
Additionally, some minor changes in the peak amplitude and in the side lobes are occurring. For the initial times of the evolution, the changes in the side lobes pulsate. Then, for longer time evolutions it is possible to observe that side lobes eventually vanish and the base of the soliton waveform thickens. In order to examine the stabilities of the self-localized one soliton profile, we present the time vs soliton power graph in Fig.~\ref{fig7}. Checking this figure, it is possible to conclude that soliton power remains bounded and has a steady-in-the-mean behavior is attained for the large times of temporal evolution. We perform a similar analysis for the value of $q=0.8$ and display the related results in Fig.~\ref{fig8}. The time vs soliton power graph behaves similarly to the graph depicted in Fig.~\ref{fig6} for the case $q=0.2$, and for the sake of brevity with better presentation purposes are not displayed.

Next, we observe the temporal dynamics of the self-localized two soliton solutions of the NLSE with a q-deformed Rosen-Morse potential and depict the results obtained for the case $q=0.8$ in Fig.~\ref{fig8} and Fig.~\ref{fig9}. Checking  Fig.~\ref{fig8} it is possible to conclude that the two-peaked soliton profile is well-preserved during temporal evolution. Additionally, the decay in the peak amplitude and development of side lobes around the main lobes of the soliton profiles are not as significant as in the case of one soliton solution. The graphs displayed in Fig.~\ref{fig9} and Fig.~\ref{fig10} clearly shows that the self-localized two soliton power remains bounded with some undulations which eventually reaches a steady-in-the-mean characteristic at larger times of evolution. Compared to its one soliton analog, the time series of the power of the two soliton solution exhibits less frequent undulations. As a result, it is possible to conclude that self-localized one and two soliton solutions of the NLSE with the q-deformed Rosen-Morse potential are stable, at least for the parameters considered.

\begin{figure}[t!]
	\begin{center}
		\hspace*{-0.7cm}
		\includegraphics[width=4.0in]{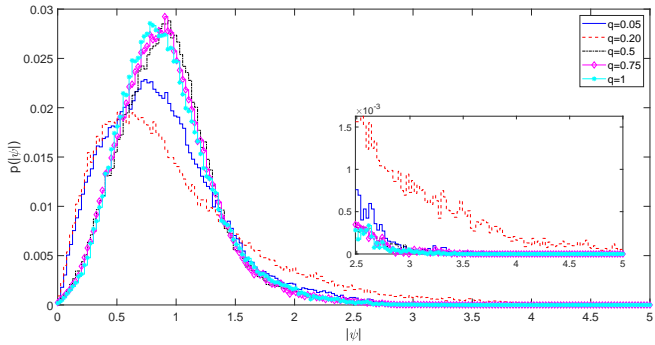}
	\end{center}
	\caption{\small Statistics of the rogue wave field for various values of q. }
	\label{fig12}
\end{figure}

One of the problems of great practical significance is the robustness and stability characteristics of these self-localized solitons under the effect of noise. It is well-known that when a monochromatic sinusoidal wave is subjected to noise within the frame of a nonlinear model, modulation instability is triggered and the monochromatic wave field turns into a chaotic polychromatic wave field exhibiting large and unexpected waves, which are known as rogue (freak) waves \cite{Kharif, Akhmediev2009b, Akhmediev2009a, Akhmediev2011, BayPRE1, BayPLA, Peregrine}. However, better terminology is the rogue oscillation if these waves are investigated within the frame of harmonic oscillators \cite{Cihan2020PhysA}. Before discussing the effects of noise on the dynamics and stability characteristics of self-localizes solitons, we first use the concept of modulation instability as a benchmark problem. With this motivation, we select an initial condition in the form
 \begin{equation}
\psi_0= e^{i m k_0 x}+ \beta a(x)
\label{eq12}
\end{equation}

\noindent where $m$ is an integer for which the value of $m=4$ is used for our calculations, $\beta$ is the amplitude of noisy perturbation which is selected as $\beta=0.2$ and a(x) is a vector including uniformly distributed random numbers in the interval of [-1,1]. We use this function as an initial condition of the Fourier spectral scheme with a $4^{th}$ order Runge-Kutta time integrator that we have implemented for the solution of the NLSE with the q-deformed Rosen-Morse potential. We observe that with this initial condition, the model we investigate exhibits rogue oscillations. A typical chaotic wave field exhibiting such rogue oscillations is displayed in Fig.~\ref{fig11}.

We also investigate the effect of the q parameter on the statistics of rogue oscillations modeled in the frame of the NLSE with the q-deformed Rosen-Morse potential and display the statistics in Fig.~\ref{fig12}. Each of the probability distribution functions (PDFs) we display in Fig.~\ref{fig12} includes approximately $10^5$ waves. These massive simulations are allowing us to reach statistically meaningful conclusions. Checking Fig.~\ref{fig12}, it is possible to conclude that rogue oscillations with amplitudes $\psi \in [0,5]$ can be observed for each q. However, the probability of observation for such a rogue oscillation is significantly higher for $q=0.2$ compared to the other values considered. Additionally, these PDFs do not behave monotonically, that is a decrease in q does not always lead to a higher probability of rogue oscillation, as Fig.~\ref{fig12} confirms.

\begin{figure}[htb!]
	\begin{center}
		\hspace*{-0.7cm}
		\includegraphics[width=4.0in]{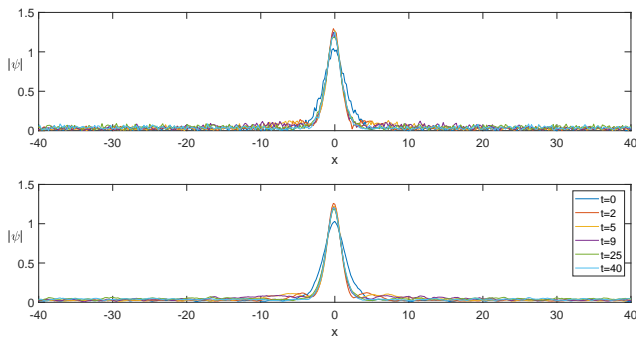}
	\end{center}
	\caption{\small Self-localized one soliton profiles at different times for $q=0.8$ a) under the effect of noise b) after they are filtered using a Savitzky-Golay filter.}
	\label{fig13}
\end{figure}

Next, we turn our attention to the dynamics and stability characteristics of the self-localized solutions of the NLSE with the q-deformed Rosen-Morse potential. When we impose similar noise terms on our self-localized solutions obtained by PM during time-stepping, we observe that they are robust under these perturbation chaotic wave fields are not generated within the frame of the model under investigation. In Fig.~\ref{fig13} subplot a), we depict the self-localized one soliton shape at different times. In the subplot b) of the same figure, we display the denoised soliton shapes at different times. Denoising is performed at the last step of temporal evolution. The denoising filter we use is a $3^{rd}$ order Savitzky-Golay filter.

\begin{figure}[htb!]
	\begin{center}
		\hspace*{-0.7cm}
		\includegraphics[width=4.0in]{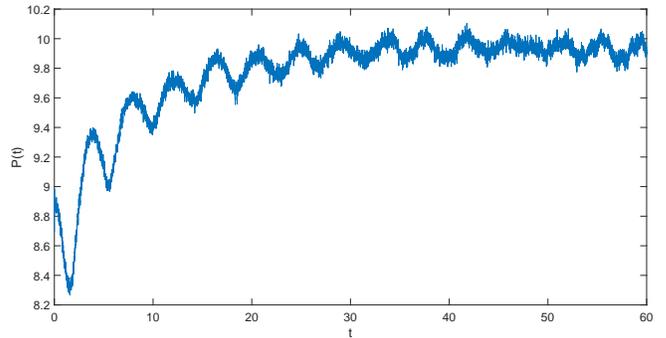}
	\end{center}
	\caption{\small Time series of the self-localized one soliton power under the effect of noise for $q=0.8$.}
	\label{fig14}
\end{figure}

In Fig.~\ref{fig14}, we display the time series of the self-localized one soliton power for $q=0.8$. As indicated in this figure, the effect of noise can be observed however power remains bounded and reaches a steady-in-the-mean behavior in the long time evolutions, similar to the noiseless case.
Checking Fig.~\ref{fig13} and Fig.~\ref{fig14}, one can conclude that self-localized solitons are robust under noisy perturbations and the dynamics of the main lobe of the soliton profile and its side lobe characteristics are similar to its noiseless analog.

\begin{figure}[b!]
\begin{center}
\hspace*{-0.7cm}
   \includegraphics[width=4.0in]{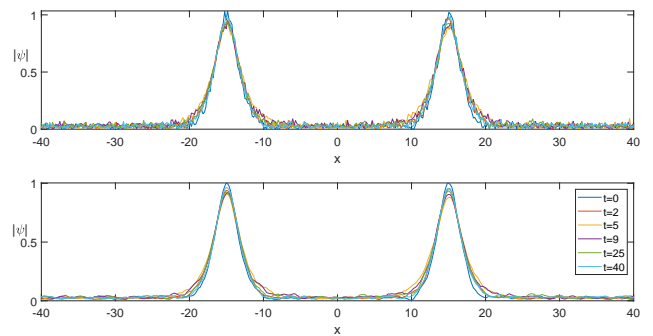}
  \end{center}
\caption{\small Self-localized two soliton profiles at different times for $q=0.8$ a) under the effect of noise b) after they are filtered using a Savitzky-Golay filter.}
  \label{fig15}
\end{figure}

Similarly, when we investigate the effect of noisy perturbation of the dynamics and stability characteristics of the self-localized two soliton solution, we obtain a similar behavior. The two soliton profiles under the effect of noise are depicted in Fig.~\ref{fig13}. Checking this figure, the same conclusions can be drawn with the one soliton case. Thus, it is possible to conclude that self-localized one and two solitons solutions of the NLSE with the q-deformed Rosen-Morse potential are robust and stable even under the effect of noise.

The soliton dynamics observed here are promising and can be used to analyze the effects of solitons on physical phenomena studied using q-deformed models, such as boson physics, thermodynamics, entanglement, and atomic physics. Our results suggest the integrability of the NLSE with the q-deformed Rosen-Morse potential and they are possibly be extended N-soliton solutions. In near future, we plan to investigate the similar characteristics of the solutions of various nonlinear systems with a parity-time symmetric Rosen-Morse potential given as $V(x)=A\text{sech}_q^2(\alpha x)+ Bi \text{tanh}_q(\alpha x)$ and with an $\epsilon$-deformed anharmonic oscillator in the form $V(x)=x^{2K}(ix)^{\epsilon}$ where $K=1,2,3,...$ and $\epsilon>0$ \cite{Bender1998, Berry380}, which in the limit $\epsilon \rightarrow 0+$ and for $K=1$, the $\epsilon$-deformed anharmonic oscillator reduces to the well-known harmonic potential $V(x)=x^2$ giving a basis to compare the results with the well-established results in the literature. Furthermore, the results and approach used in this paper can also be used to investigate many different nonlinear phenomena in atomic physics, optics, hydrodynamics, traffic flow theory, and Bose-Einstein condensation, just to name a few.

\section{Conclusion}
In this paper, we numerically investigated the characteristics and stabilities of the self-localized solutions of the NLSE with a q-deformed Rosen-Morse potential. Using the Petviashvili method (PM), we showed that self-localized one and two soliton solutions of the NLSE with such a potential term do exist. The q-deformed Rosen-Morse potential leads to a scaling of the soliton amplitude as well as a spatial shift in the location of the peak of the soliton. These solitons also satisfy the Vakhitov-Kolokolov condition which is one of the necessary conditions for the linear stability. We have also studied the nonlinear stabilities of the aforementioned solitons using a time-stepping algorithm. The numerical algorithm we implemented for this purpose was a Fourier spectral algorithm with a $4^{th}$ order Runge-Kutta time integrator. We observed that during the temporal evolution main lobe of the soliton is well-preserved with some minor changes in peak amplitude and minor side lobe changes occurring in a pulsating manner for the initial times of the time-stepping. For longer time evolutions we observed that side lobes vanish and the base of the soliton profile thickens. In order to examine the stabilities of these solitons, we also presented time vs soliton power characteristics. As a result, we observed that self-localized one and two soliton solutions of the NLSE with a q-deformed Rosen-Morse potential are stable. We have also investigated the robustness of the aforementioned solitons against noise. With this motivation, we first observed that noise imposed on a monochromatic sinusoidal wave turns the monochromatic wave field into a chaotic one which exhibits many peaks in the form of rogue oscillations. We observed that the effect of q-deformed Rosen-Morse potential on the statistics of these rogue oscillations having peaks in the interval of $|\psi| \in [0,5]$ are not behaving monotonically. That is, depending on the value of the q-deformation parameter, an increase in the value of q may lead to a higher or lower probability of occurrence of rogue oscillations. However, when we investigated the noisy dynamics of the self-localized solitons, we observed that they are robust against white noise perturbations and well-defined soliton shapes are well-preserved and the soliton power remains bounded. We have also discussed a possible way of denoising the solitons using a Savitzky-Golay filter. To sum, we showed that self-localized one and two soliton solutions of the NLSE with a q-deformed Rosen-Morse potential are stable even if they are perturbed by white noise.

The results and approach presented in this paper can be extended and applied to investigate many different phenomena in nonlinear physics. Our results can be immediately used to investigate the effects of soliton solutions on physical phenomena studied using q-deformed models, including but are not limited to boson physics, thermodynamics, entanglement, and atomic behavior. It is possible to follow a similar route to analyze the one, two or N-soliton solutions of the NLSE or NLSE type dynamic equations having different potentials including but are not limited to Eckart, parity-time symmetric Rosen-Morse, $\epsilon$-deformed an-harmonic, Coulomb or Yukawa potentials, just to name a few. Such an analysis would bring significant improvements and new insights into atomic and subatomic physics. Additionally, one can extend our analysis to higher dimensional uncoupled or coupled systems. It is also possible to investigate the interaction of the solitons of such systems during time evolution. Analysis of such systems would also be very beneficial in understanding different nonlinear phenomena in optics, hydrodynamics, and Bose-Einstein condensation.

\section{Acknowledgment}
\noindent F.O. acknowledges the financial support of Tokyo International University Personal Research Fund.

\end{document}